\documentclass{aa}  

\usepackage{graphicx}
\usepackage{graphicx}
\usepackage{lscape}
\usepackage{longtable}
\usepackage{natbib}
\usepackage{color}
\usepackage{array} 
\bibpunct{(}{)}{;}{a}{}{,} 

\usepackage{txfonts}
\begin{document}

   \title{Close-in planets around giant stars}

   \subtitle{Lack of hot-Jupiters and prevalence of multi-planetary systems}

   \author{J. Lillo-Box\inst{1}, D. Barrado\inst{2}, A. C. M. Correia\inst{3,4} 
          }

  \institute{European Southern Observatory (ESO), Alonso de Cordova 3107, Vitacura, Casilla 19001, Santiago de Chile (Chile)\\
              \email{jlillobox@eso.org}\and
Depto. de Astrof\'isica, Centro de Astrobiolog\'ia (CSIC-INTA), ESAC campus 28691 Villanueva de la Ca\~nada (Madrid), Spain \and
CIDMA, Departamento de F\'isica, Universidade de Aveiro, Campus de Santiago, 3810-193 Aveiro, Portugal  \and
ASD, IMCCE-CNRS UMR8028, Observatoire de Paris, 77, Av. Denfert-Rochereau, 75014 Paris, France
            }

   \date{Accepted on March 11th, 2016}

\titlerunning{Close-in planets around giant stars: Lack of hot-Jupiters and prevalence of multi-planetary systems}
\authorrunning{Lillo-Box et al.}

 
  \abstract
   {Extrasolar planets abound in almost any possible configuration. However, until five years ago, there was a lack of planets orbiting closer than 0.5 au to giant or subgiant stars. Since then, recent detections have started to populated this regime by confirming 13 planetary systems. We discuss the properties of these systems in terms of their formation and evolution off the main sequence. Interestingly, we find that $70.0\pm6.6$\% of the planets in this regime are inner components of multiplanetary systems. This value is 4.2$\sigma$ higher than for main-sequence hosts, which we find to be $42.4\pm0.1$\%. {The properties of the known planets seem to indicate that the closest-in planets ($a<0.06$~au) to main-sequence stars are massive (i.e., hot Jupiters) and isolated and that they are subsequently engulfed by their host as it evolves to the red giant branch, leaving only the predominant population of multiplanetary systems in orbits $0.06<a<0.5$~au. We discuss the implications of this emerging observational trend in the context of formation and evolution of hot Jupiters.}}
      

   \keywords{Planets and satellites: gaseous planets, planet-star interactions, dynamical evolution and stability   }

   \maketitle
%

\section{Introduction}

The large crop of extrasolar planets discovered so far offers a new possibility of studying the properties of these systems from a statistical point of view \citep[e.g.,][]{batalha14}. 
This bounty of planetary systems and their widely diverse properties allow us to start answering the question of how these bodies form and evolve from an observational point of view. However, while hundreds of planets have been found around solar-mass stars, there is still a scarcity of planets orbiting more massive hosts. Some of these hosts have already left the main sequence, now being in the giant or subgiant phase.  In consequence, looking for planets around these more evolved stars (with much sharper absorption lines) can help to better constrain the demography of planets around early-type stars and so probing the efficiency of planet formation mechanisms for the different mass ranges of the host star. The discovery of planets around K and G giants and subgiants is therefore crucial for planet formation theories. 

From the observational point of view, there is a dearth of planets with short periods around stars ascending the red giant branch (RGB, \citealt{johnson07}). The reason could be twofold. On one hand, it could be explained by a scarcity of close-in planets around early-type main-sequence stars ($M_{\star} > 1.2 M_{\odot}$). This scarcity has been hypothesized to be related to different migration mechanisms for planets around stars of different masses (Udry et al., 2003), owing to the shorter dissipation timescales of the protoplanetary disks of these stars, which prevents the formed planets from migrating to close-in orbits \citep[see, e.g.,][]{burkert07, currie09}. 
On the other hand, the paucity of planets around these evolved stars has been considered by theoretical studies as evidence of the planet engulfment or disruption even in the first stages of the evolution off the main sequence of their parents. \cite{villaver09} calculated how tidal interactions in the subgiant and giant stages can lead to the final engulfment of the close-in planets and how this process is more efficient for more massive planets. Likewise,  \cite{villaver14} have analyzed the effects of the evolution of the planet's orbital eccentricity, mass-loss rate, and planetary mass on the survivability of planets orbiting massive stars ($M_{\star} > 1.5 M_{\odot}$). They conclude that the planet mass is a key parameter for the engulfment during the subgiant phase, with the more massive planets more likely falling into the stellar envelope during this phase (for the same initial orbital parameters). Also, planets located at 2-3 $R_{\star}$, when the star begins to leave the main sequence, may suffer orbital decay from the influence of stellar tides. Any planet closer than this orbital distance may be engulfed by the star in the subgiant phase. However, these results are based on a limited sample of confirmed exoplanets around RGB stars. Therefore the detection of  close-in planets around post main-sequence stars is crucial for constraining theoretical models of planet engulfment. Several long-term projects have therefore focused on finding these planets (e.g., TAPAS: \citealt{niedzielski15}; EXPRESS: \citealt{jones11}).

In this paper, we analyze the growing sample of close-in planets around giant stars {from an observational point of view}. {We analyze their properties and multiplicity and compare them to planets around main-sequence stars, trying to determine the evolution of these properties. }{We discuss the possible explanations for the observed trend in the context of the formation and evolution of hot-Jupiter planets and the consequences of their inward migration.}

\section{Properties of  the known close-in planets around giant stars}
\label{sec:properties}

\begin{figure}[hbtp]
\includegraphics[width=0.5\textwidth]{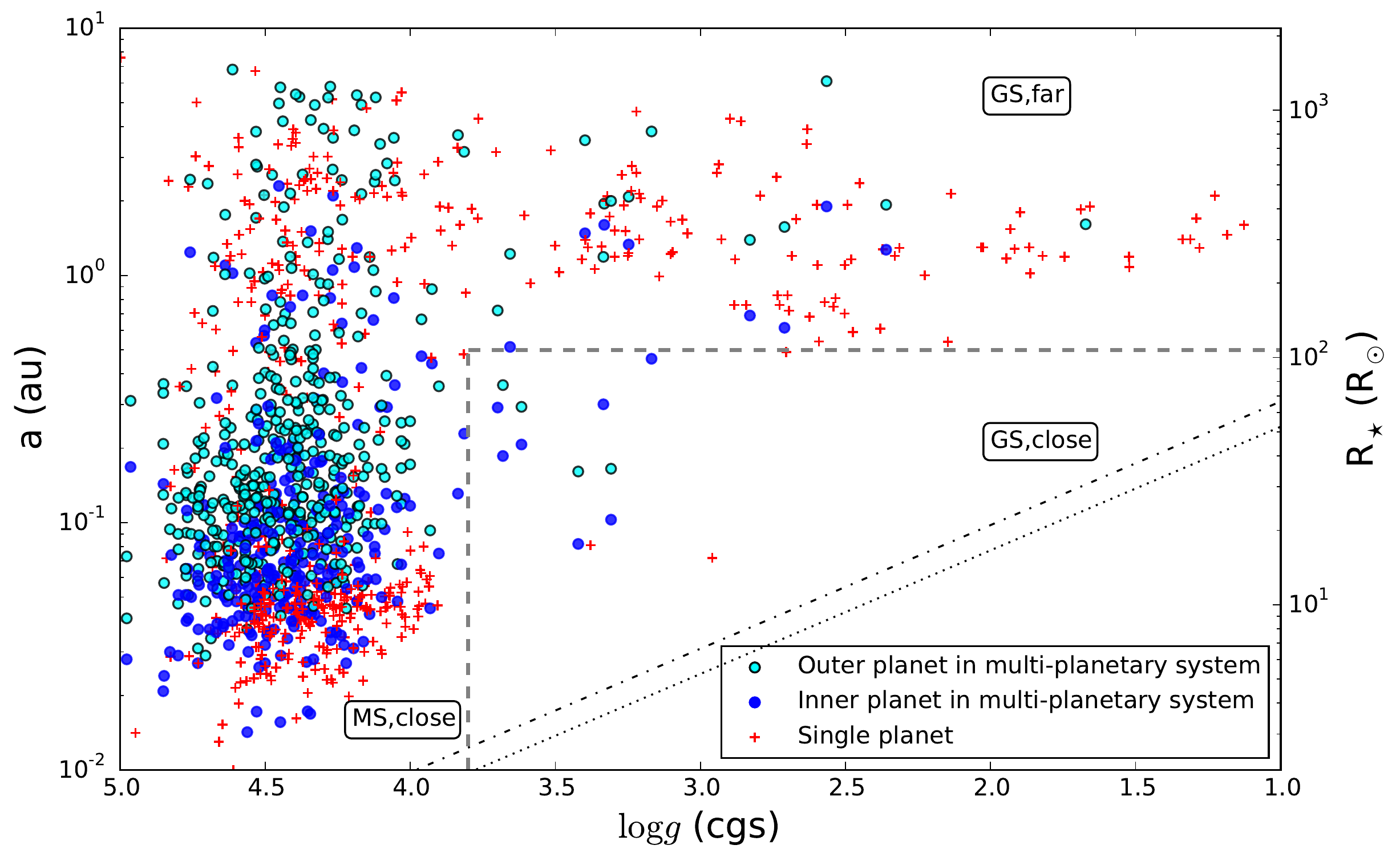}
	\caption{Semi-major axis of known planets against their host surface gravity. We include single planets (red plus symbols) and multiplanetary systems (circles) with the inner component being the dark blue and the outer planet(s) in light blue. {For reference, we also plot the stellar radius of the host star at every evolutionary stage (i.e., for the different surface gravities) for two masses that comprise the range of masses of the hosts analyzed in this paper, $1~M_{\odot}$(dotted) and $1.6~M_{\odot}$(dash-dotted).} }	
	\label{fig:logg_sma}
\end{figure}

From the sample of confirmed or validated extrasolar planets, we have selected those with close-in orbits around stars in the subgiant phase or ascending the RGB. We consider planets in this regime as those revolving in orbits closer than $a<0.5$~au around host stars with $\log{g}<3.8$. The limit in semi-major axis was chosen as the apparent limit closer than which no planets have been found until recent years. The limit in surface gravity was chosen to select both giant and subgiant stars. In Fig.~\ref{fig:logg_sma}, we show the location of all known planets in the $[a,\log{g}]$ parameter space. In total, 13 systems have been identified so far in the mentioned regime (including uncertainties):  Kepler-56, Kepler-91, Kepler-108, Kepler-278, Kepler-368, Kepler-391, Kepler-432,  8\,Umi, 70\,Vir, HD\,11964, HD\,38529 , HD\,102956, and HIP\,67851.  In Table~\ref{tab:properties} we present the main properties of the planets and host stars in these systems. We include all systems that lie inside the above-mentioned boundaries within their $3\sigma$ uncertainties. In total, the sample is composed of four su-giant  hosts and nine stars already ascending the RGB.

\section{Results}
\label{sec:results}

We can see from Fig.~\ref{fig:logg_sma} that most of the planets with $\log{g}<3.8$ and $a<0.5$~au are multiplanetary systems with at least the inner component in this regime. In four cases, no additional planets are reported by the discovery papers. These are Kepler-91, HD\,102956, 8\,Umi, and 70\,Vir. In this section we calculate the actual rate (and uncertainty) of multiplanetary systems in this regime and analyze the RV observations of the single systems to put constraints on possibly undetected outer planets.

\begin{figure*}[hbtp]
\includegraphics[width=1.0\textwidth]{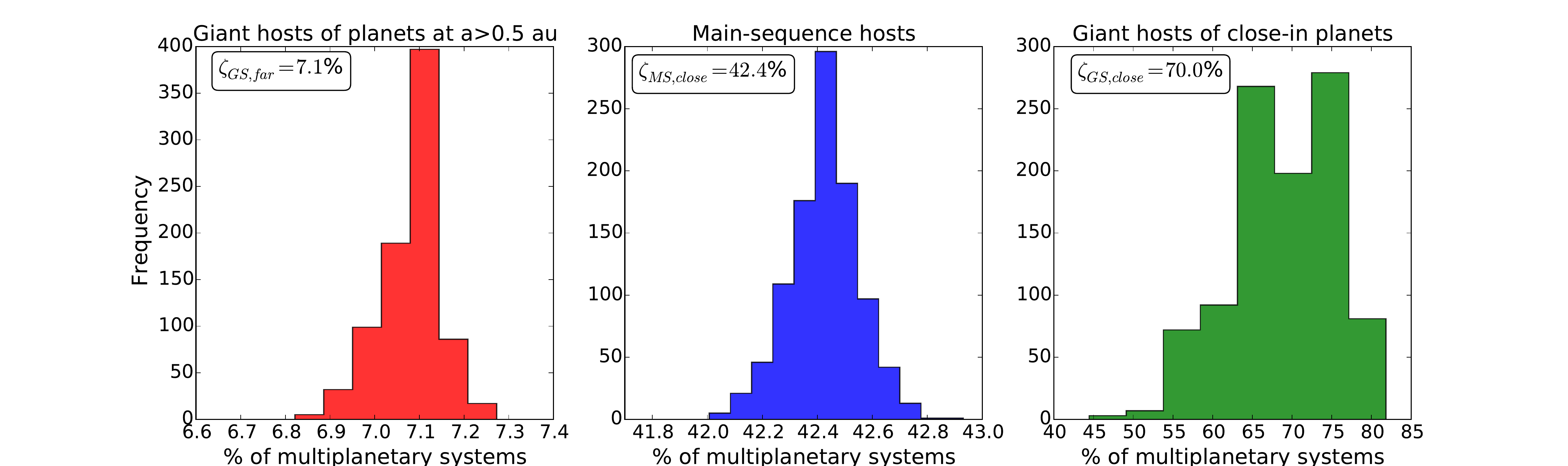}	
\caption{Distribution of the percentage of multiplanetary systems with the inner component in three different regimes: beyond 0.5 au around giant stars (left panel), closer than 0.5 au around main-sequence stars (middle panel), and closer than 0.5 au around giant stars (right panel).}	
	\label{fig:distributions}
\end{figure*}

We retrieved the host masses and radii from the Exoplanet Catalogue\footnote{\url{exoplanet.eu}} and the NASA Exoplanet Archive\footnote{\url{exoplanetarchive.ipac.caltech.edu}}. The surface gravity and its uncertainty is then computed from those values. The semi-major axis was also retrieved from the same source. We divide the $[\log{g},a]$ parameter space into three regimes (see Fig.~\ref{fig:logg_sma}): close-in planets around main-sequence stars (MS$_{\rm close}$, $\log{g}>3.8$ and $a<0.5$~au), close-in planets around giant stars (GS$_{\rm close}$, $\log{(g)}<3.8$ and $a<0.5$~au), and far away planets around giant stars (GS$_{\rm far}$, $\log{g}<3.8$ and $a>0.5$~au). We bootstrap the values of all planets by considering Gaussian distribution of probability with standard deviation equal to the uncertainty in the parameters. In each step, we count the number of planets in each particular regime that are inner components in multiplanetary systems with respect to the total number of systems in that regime ($\zeta$). We ran $10^3$ steps and obtained the distribution of $\zeta$ for each regime (see Fig.~\ref{fig:distributions}). 

The median and standard deviation of the distribution of the bootstrapped populations for each regime are kept. These values represent the ratio of inner planets in multi-planetary systems in each of the three regimes. We find that $\zeta_{\rm MS,close}=42.4\pm0.1$\% of the planetary systems hosted by main sequence stars with at least one component closer than 0.5 au are multiplanetary systems. Regarding systems in orbits farther than 0.5 au, we find that $\zeta_{\rm GS,far}=7.1\pm0.1$\% of the systems in this regime are multiple systems with their inner component farther than 0.5 au. Finally, we find that in $\zeta_{\rm GS,close}=70.0\pm6.6$\% of the systems with the closer planet more interior than 0.5 au, this planet is the inner component of a multi-planetary system. According to this, the ratio of close-in planets that are inner components in multiplanetary systems is significantly greater ($>4.2\sigma$) in the case of giant stars than it is during the main-sequence stage. The implications of this significant difference are discussed in Sect.~\ref{sec:discussion}.

We must note that we are aware that these numbers may suffer from some observational biases (specially $\zeta_{\rm GS,far}$) since, {for instance, detecting outer components to planets with $a>0.5$~au around giant stars by} both radial velocity and transit methods is much more difficult or, at the least, less likely. Also, detecting planets with the transit method is more difficult in the case of giants owing to the contrast radius ratio. These biases have not been taken into account in this calculation. However, it is important to point out that both biases would favor an increase of $\zeta_{\rm GS,close}$ over  $\zeta_{\rm MS,close}$.

\begin{figure*}[htbp]
\includegraphics[width=0.33\textwidth]{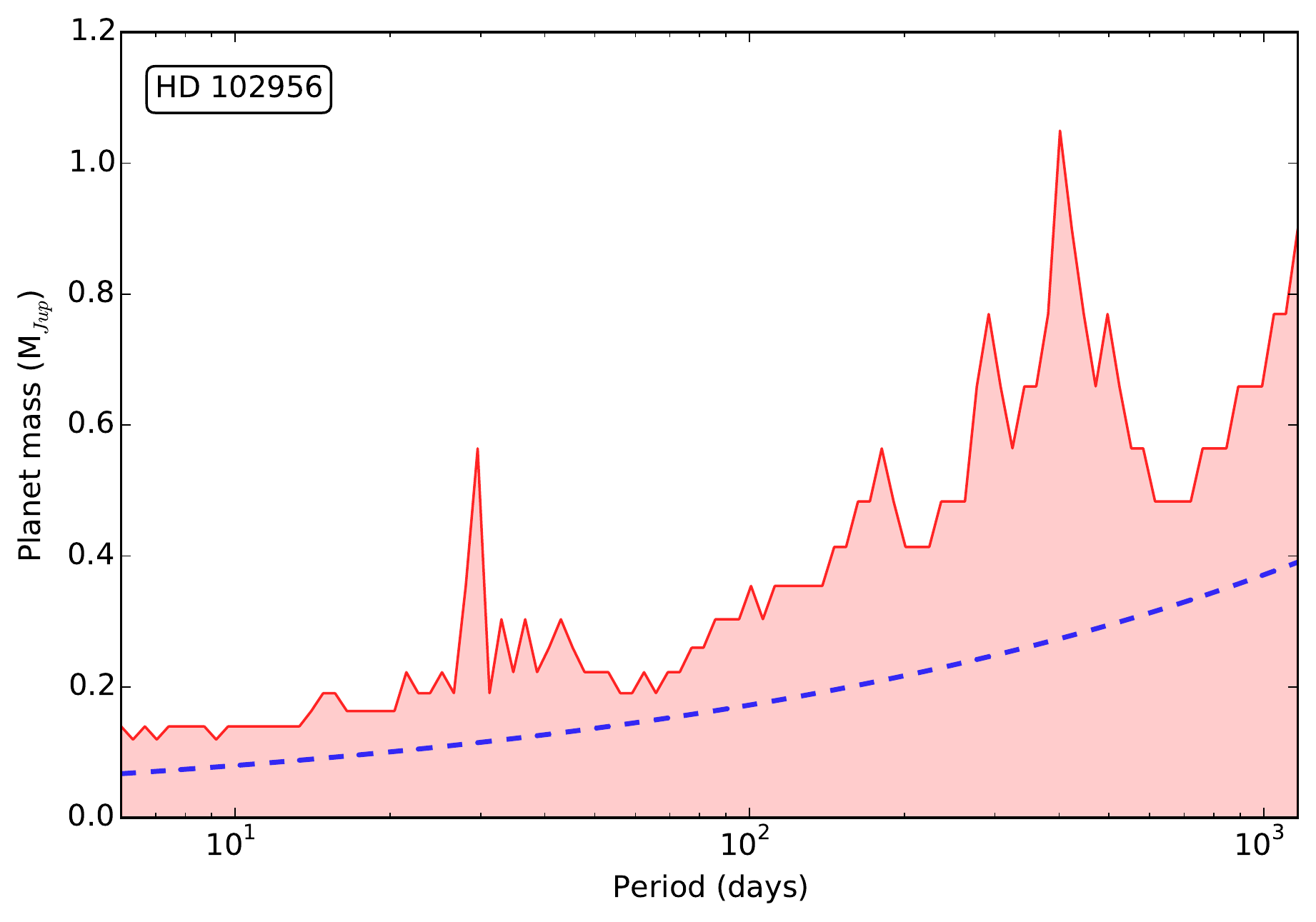}
\includegraphics[width=0.33\textwidth]{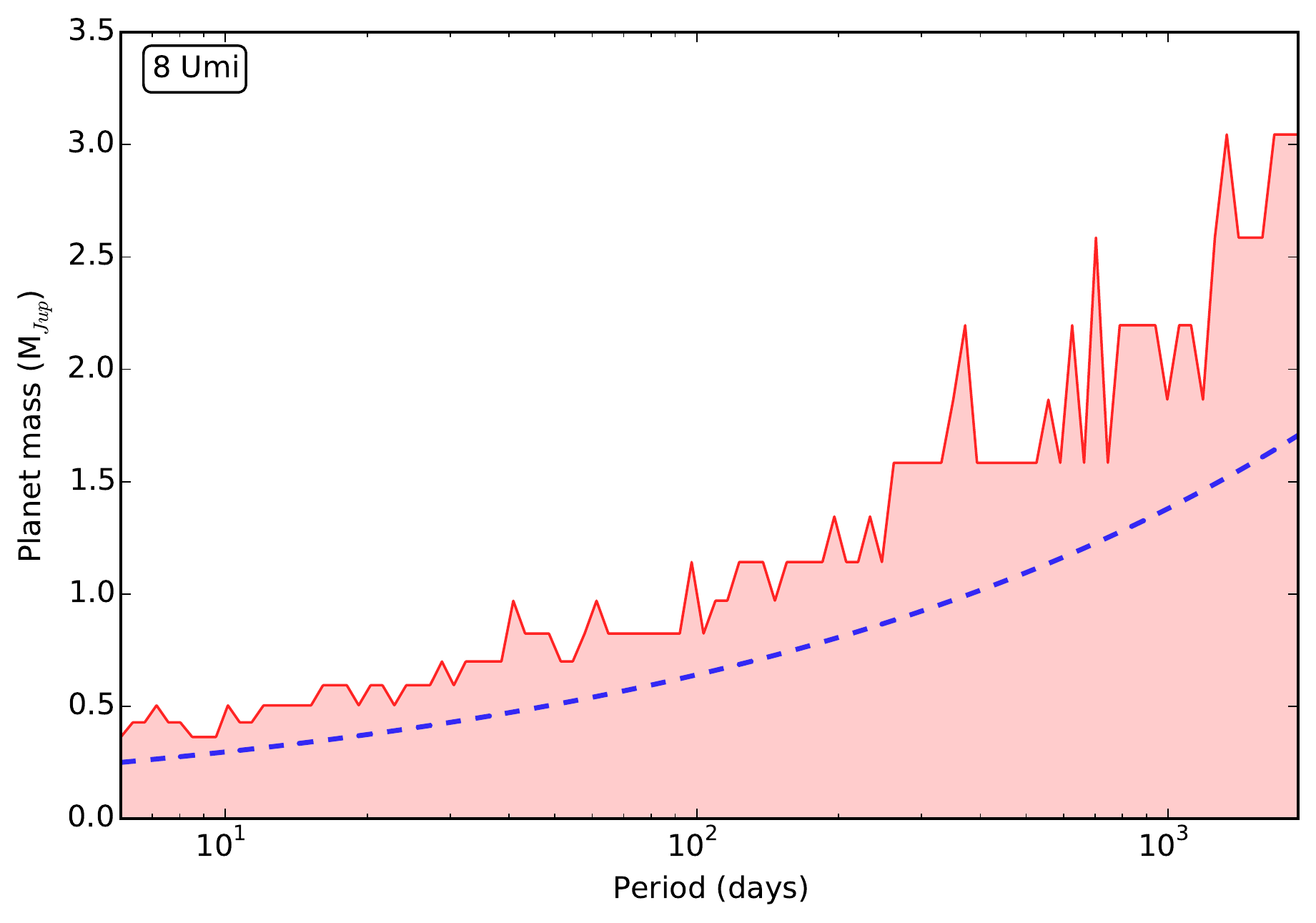}
\includegraphics[width=0.33\textwidth]{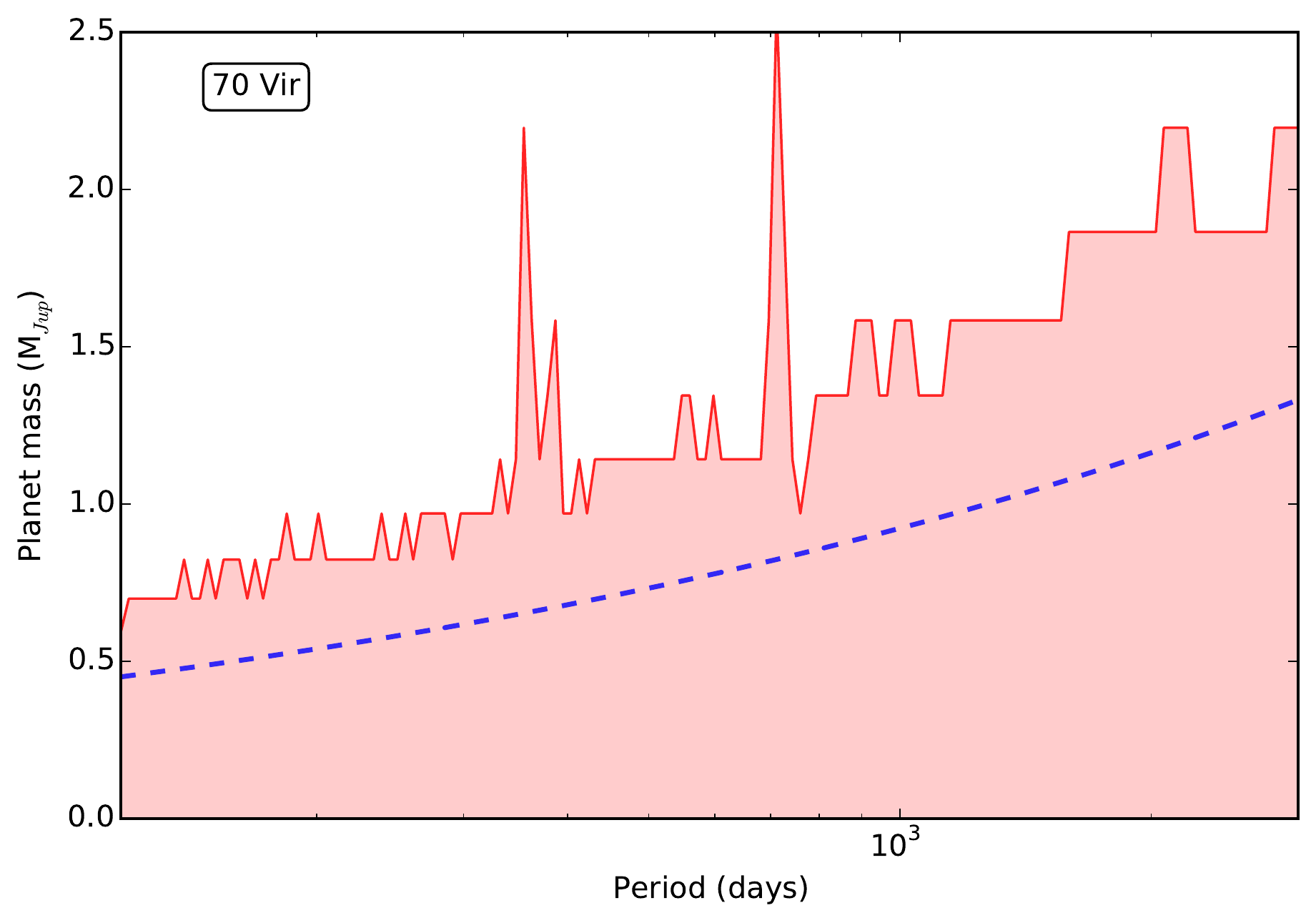}
	\caption{Parameter space for the detectability of longer period planets in the systems HD\,102956 (left), 8 Umi (middle), and 70 Vir (right), assuming circular orbits. Planets located in the red shaded region are not detectable by the radial velocity data obtained by \cite{johnson10}, \cite{lee15}, and \cite{naef03}, respectively. The blue dashed line indicates the planet mass that would produce a radial velocity amplitude equal to the standard deviation of the radial velocity data (once the signal of the inner planet has been removed). See details on the calculation in Sect.~\ref{sec:results}.}	
	\label{fig:minimumMp}
\end{figure*}

Among the four planetary systems in close-in orbits around giant stars found to host only one planet, we have investigated the mass-period parameter space yet unexplored by the discovery RV data to put constraints on additional undetected planet.
We have used the RV data published in the discovery papers of HD\,102956 (22 epochs in a 1164-days timespan; \citealt{johnson10}), 8 Ursa Minor (21 measurements in a 1888-day timespan; \citealt{lee15}), and 70 Vir (\citealt{naef03}) to check for the limitations of these observations for detecting additional planets in the system. We used these data and the derived parameters in the discovery works to remove the contribution of the known planet and look for a second planet in the residuals. 

This was done by following the prescriptions in \cite{lagrange09}, by simulating a planet signal  in a circular orbit with the corresponding $[m_p,P_{\rm orb}]$ and observed at the same dates, and by introducing the typical noise for the particular instrument used in the discovery of the inner planet. The results of the simulation are provided in Fig.~\ref{fig:minimumMp}, where we set the properties of the planets that could have been missed by the discovery papers. Regarding Kepler-91 \citep{lillo-box14,lillo-box14c}, subsequent follow-up observations after its confirmation pointed out the possibility that additional bodies exist by detecting a possible radial velocity drift \citep{sato15} or by detecting additional dips in the \emph{Kepler} phase-folded light curve.

In these four cases, the possibility of an additional outer planet is not discarded among the presented limits, but we count them as single for the purposes of this work. 

\section{Discussion and conclusions}
\label{sec:discussion}

The results presented in the previous section indicate that {(according to observations)} the large majority of planetary systems in close-in orbits around giant stars are multiple. We have found that the ratio of these systems against the total number of systems is significantly higher in giant and subgiant stars than in main-sequence hosts. Even though we are in the low number statistics, there seems to be an increasing sample of multiplanetary systems around evolved stars with their inner component in an orbit closer than 0.5 au. We thus may wonder about the reasons for this higher rate against single planets {and about how this is related to the evolution of planetary systems.} 

\begin{figure*}[htbp]
\centering
\includegraphics[width=0.45\textwidth]{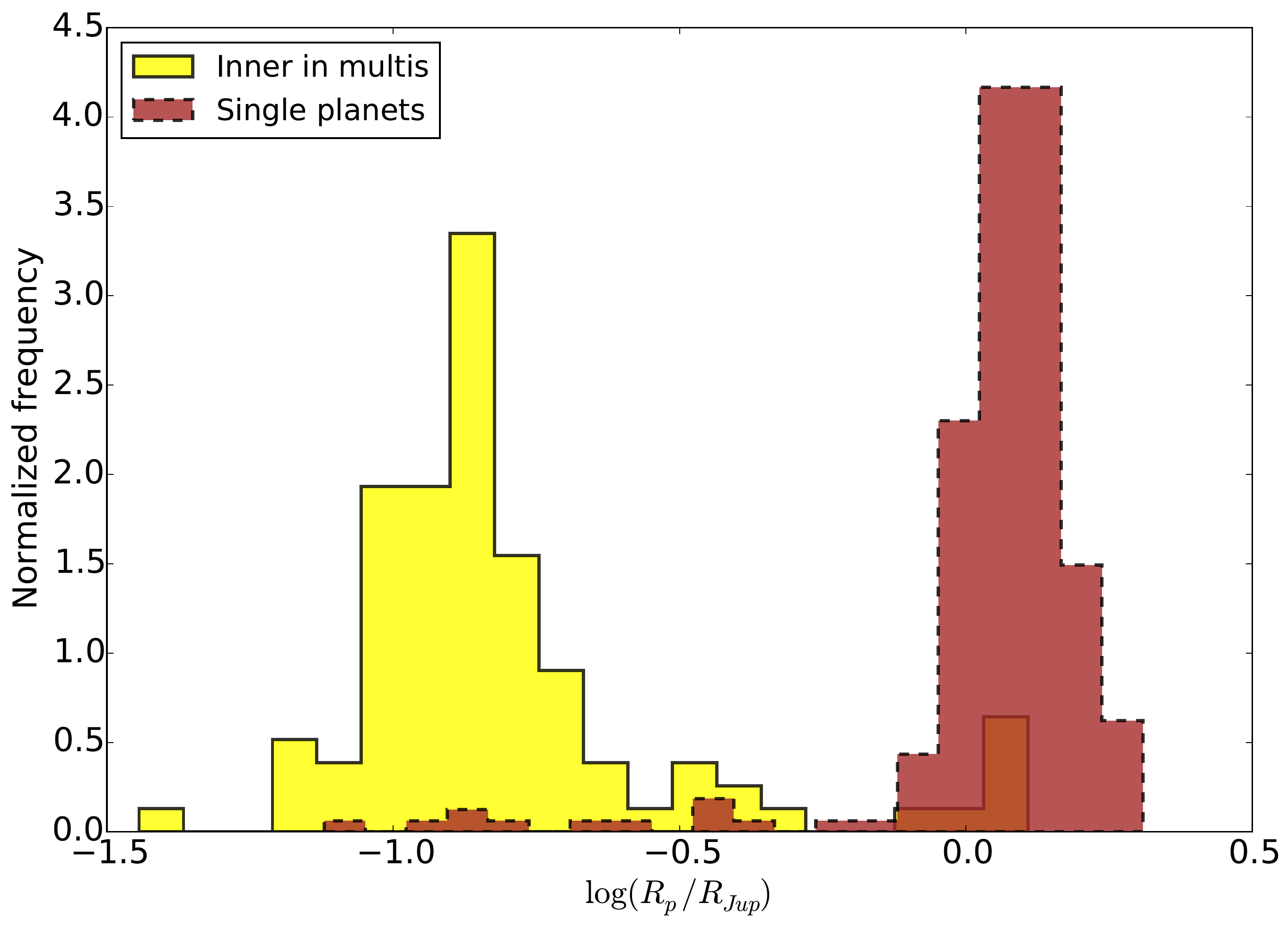}
\includegraphics[width=0.45\textwidth]{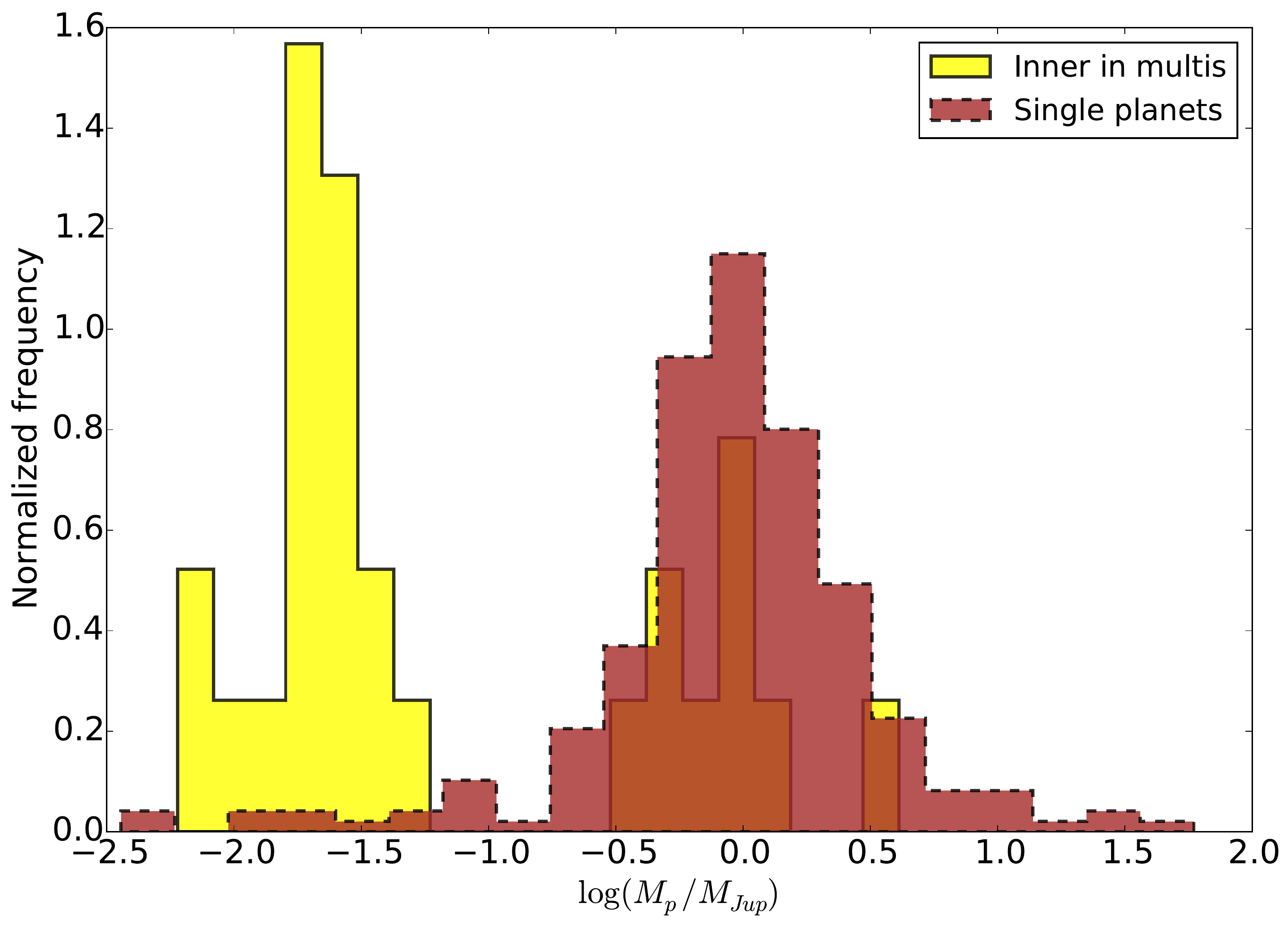}
	\caption{{Planetary mass distribution for planets closer than 0.06 au to main-sequence stars. The sample is divided into those systems where only one planet has been detected in the system (brown histogram with dashed line) and planets that are inner components of multiplanetary systems (yellow histogram with solid line). }}	
	\label{fig:closein}
\end{figure*}

It has been theorized that giant planets formed in the outer regions of the disk are capable of destroying or swallowing smaller planets in their inward migration to the inner parts of the system \citep[e.g., ][]{mustill15}. This could explain the high concentration of single planets in orbits closer than 0.06 au  around main-sequence stars (see Fig.~\ref{fig:logg_sma}). 
{Indeed, 68.1\% of the planets found in this regime are found to be single planets, while inner components of multiplanetary systems represent the 31.0\% of the systems. By contrast, planets at 0.06-0.5 au are mostly multiple systems with at least the inner component in this range. According to the sample of known planets used in this paper, we find that 64.5\% of the systems in this regime are inner planets in multiplanetary systems, while just 35.5\% are single-planet systems. \\ By focusing on the properties of the extremely close-in planets ($a<0.06$~au) around main-sequence stars, we see a clear segregation, with the single planets being Jupiter-like and the inner components of multiplanetary systems being small (rocky) planets.  This segregation is clear in Fig.~\ref{fig:closein}, where we show the mass and radius distribution of extremely close-in planets around main-sequence stars for the sample of single planetary systems and inner planets in multiplanetary systems. From this reasoning, it is clear that during the main-sequence stage, planets in these extremely close-in orbits are mainly single Jupiter-like planets (i.e., hot Jupiters). This observational result is in good agreement with the idea that the migration of massive planets from the outer regions of the disk tends to destroy any already formed rocky planets in the inner regions. \\ It is important to point out that this destruction of the smaller planets may not always take place as shown in the case of the WASP-47 system, where a Neptune-sized outer planet and a super-Earth inner companion  to a hot-Jupiter planet were found by \cite{becker15}. This discovery highlights that several effects may be taking place during this process.}

{The previously summarized conclusions are relevant for the results of this work since we have not found extremely close-in planets around giant  or subgiant stars. However, we find planets in the 0.06-0.5 au regime around these evolved stars. And, more importantly, we find that they are mostly inner components in multiplanetary systems in a fraction of $70.0\pm6.6$~\%, which is compatible with the 64.5\% previously mentioned for the main-sequence systems in the same regime. The observational conclusion that can be extracted from these numbers is that once the star evolves off the main-sequence, the closest planets (usually massive and single) are swallowed by their host, leaving only the population of multiplanetary systems in orbits beyond 0.06 au. Numerical simulations by \cite{frewen16} have shown that this could indeed explain the deficiency of warm Jupiters in stars larger than two solar radii. They hypothesize that these gaseous planets with periods of 10-100 days may migrate owing to Kozai-Lidov oscillations with a subsequent decay via tidal interactions with the star. In the current sample, the single planets Kepler-91\,b and HD\,102956\,b are indeed single hot Jupiters in a very close-in orbit ($a<0.1$~au), so they seem to have (by now) survived this phase.}

{According to \cite{villaver14}, more massive planets are more easily and more rapidly engulfed during the evolution of their host star off the main sequence. It is thus clear from observations that hot Jupiters should practically disappear at these stages. Low-mass planets, however, could survive at least during the first stages of the evolution of the star. But, their detection is much more challenging given the small radius ratio compared to the giant planets (i.e., it makes it difficult to detect their transits) and the low-mass ratios combined with stellar pulsations during this evolutionary stage (what difficult the detection by the radial velocity technique). However, small planets should be there, and future instrumentation (e.g., ESPRESSO/VLT) will be capable of detecting them, filling the new gap of extremely close-in planets around evolved stars.}

{There are other possibilities for explaining the prevalence of close-in multiplanetary systems around giant stars. For instance, gravitational interactions between the planets can play an important role during the subgiant phase, although they tend to destabilize the system \citep{voyatzis13,veras13,mustill14}. However, mean motion resonances (MMR) can help the inner orbit to migrate outward and thus escape being engulfed by the star, as in the formation of the solar system \citep[e.g., ][]{levison03,dangelo12}. Among multiplanetary systems with the inner component being closer than 0.5 au, we checked for possible MMR. Only Kepler-56 has a near MMR of 2:1, as pointed out in the discovery paper \citep{steffen13}. In two other cases, Kepler-108 and Kepler-432, the planets are close to MMR 4:1 and 8:1. The remaining systems do not present any resonant or near-resonant periodicities. Although in a small sample, it is therefore more difficult to explain the major presence of multiplanetary systems than single planets around subgiant and giant stars only in terms of gravitational interactions between the planets in the system.}

Finally, we must note that the results published here are purely observational. They can feed theoretical studies of the evolution of planetary systems across the different stellar stages to explain the planet-star interactions once the star leaves the main sequence. However, these studies are beyond the scope of this observational work, which aims to highlight the need for detecting more close-in planets around giant stars and to provide some hints for theoretical analysis.


\begin{acknowledgements}
      This research was partially funded by Spanish grant AYA2012-38897-C02-01. This work was co-funded under the Marie Curie Actions of the European Commission (FP7-COFUND). A.C. acknowledges support from CIDMA strategic project UID/MAT/04106/2013.
\end{acknowledgements}


\bibliographystyle{aa} 
\bibliography{biblio2} 

\begin{table*}
\setlength{\extrarowheight}{5pt}
\scriptsize
\caption{Properties of the 13 close-in planets around giant stars. Data from discovery papers.}             
\label{tab:properties}      
\centering          
\begin{tabular}{r|ccc|ccccc}     
\hline\hline       

System  &  \multicolumn{3}{c|}{Host star}  &  \multicolumn{5}{c}{Planet}   \\ \hline
        &  $M_{\star}$ & $R_{\star}$ & $\log{g}$ & ID & $M_{p}$ & $R_{p}$ & P & a  \\
        &  $(M_{\odot})$ & $(R_{\odot})$ & (cgs) &    & $(M_{\rm Jup})$ & $(R_{\rm Jup})$ & (days) & (au)  \\ \hline

\multicolumn{8}{l}{\textbf{Known multi-planetary systems}}\\ \hline

\object{HD\,11964} & $1.125$ & $2.18\pm0.29$ & $3.81\pm0.12$  & b & $0.622\pm0.056$ & -  & $1945 \pm 26$ & $3.16\pm0.19$  \\
 & & &  & c & $0.079\pm0.010$ & - & $37.910 \pm 0.041 $ & $0.229\pm0.013$  \\


\object{HD\,38529} & $1.48\pm0.05$ & $2.44\pm0.22$ & $3.835\pm0.080$  & b & $0.78$ & -  & $14.3104 \pm 0.0002$ & $0.131\pm0.002$  \\
 & & &  & c & $17.7^{+1.7}_{-1.4}$ & - & $2134.76 \pm 0.4 $ & $3.695\pm0.043$  \\

\object{HIP\,67851} & $1.63\pm0.22$ & $5.52\pm0.44$ & $3.168\pm0.091$  & b & $1.38\pm0.15$ & -  & $88.9 \pm 0.1$ & $0.46\pm0.02$  \\
 & & &  & c & $5.98\pm0.76$ & - & $2131.8 \pm 88.3 $ & $3.82\pm0.23$  \\

\object{Kepler-56} & $1.32\pm0.13$ & $4.23\pm0.15$ & $3.306\pm0.045$  & b & $0.070\pm 0.012$ & $0.58\pm0.03$ & $10.5016 \pm 0.0011$ & $0.1028\pm0.0037$  \\
 & & &  & c & $0.569\pm 0.070$ & $0.88\pm0.04$ & $21.40239 \pm 0.00062$ & $0.1652\pm0.0059$  \\
 & & &  & d & $3.3$ & - &  - & 2  \\

\object{Kepler-108} & $1.377\pm0.089$ & $2.19\pm0.12$ & $3.9\pm0.035$  & b & - & $0.772\pm0.043$  & $49.183921 \pm 0.000054$ & $0.292$  \\
 & & &  & c & - & $0.730\pm0.045$ & $190.323494 \pm 0.00099 $ & $0.721$  \\

\object{Kepler-278} & $1.298\pm0.076$ & $2.94\pm0.07$ & $3.617\pm0.033$  & b & - & $0.363\pm0.014$  & $30.160546 \pm 0.000311$ & $0.207$  \\
 & & &  & c & - & $0.320\pm0.038$ & $51.078775 \pm 0.00089$ & $0.294$  \\

\object{Kepler-391} & $1.32\pm0.32$ & $2.6\pm1.2$ & $3.74\pm0.59$  & b & - & $0.285\pm0.070$  & $7.416755 \pm 0.000129$ & $0.082$  \\
 & & &  & c & - & $0.316\pm0.077$ & $20.48544 \pm 0.00032 $ & $0.161$  \\

\object{Kepler-368} & $0.77\pm0.12$ & $2.02\pm0.56$ & $3.68\pm0.25$  & b & - & $0.291\pm0.081$  & $26.84768 \pm 0.00033$ & $0.186$  \\
 & & &  & c & - & $0.346\pm0.096$ & $72.379334 \pm 0.00137 $ & $0.36$  \\

\object{Kepler-432} & $1.35\pm0.10$ & $4.15\pm.12$ & $3.331\pm0.008$  & b & $5.41^{+0.19}_{-0.30}$ & $1.450\pm0.039$  & $52.501134 \pm 0.00011$ & $0.301^{+0.011}_{-0.016}$  \\
 & & &  & c & $2.43\pm0.23$ & -  & $406.2^{+3.9}_{-2.5}  $ & $1.178^{+0.063}_{-0.042}$  \\ \hline

\multicolumn{8}{l}{\textbf{Assumed single planetary systems}}\\ \hline

\object{Kepler-91} & $1.31\pm0.10$ & $6.30\pm0.16$ & $2.953\pm0.007$  & b & $0.88^{+0.17}_{-0.33}$ & $1.384^{+0.011}_{-0.054}$ & $6.246580 \pm 0.000082$ & $0.072^{+0.002}_{-0.007}$  \\

\object{HD\,102956} & $1.68\pm0.11$ & $4.4\pm0.1$ & $3.378\pm0.035$  & b & $0.96\pm0.05$ & -  & $6.4950 \pm 0.0004$ & $0.081\pm0.002$  \\

\object{8 Umi} & $1.8\pm0.1$ & $9.9\pm0.4$ & $2.704\pm0.043$  & b & $1.5\pm0.2$ & -  & $93.4 \pm 4.5$ & $0.49\pm0.03$  \\

\object{70 Vir} & $1.09\pm0.02$ & $1.94\pm0.05$ & $3.816\pm0.024$  & b & $7.40\pm0.02$ & -  & $116.6926 \pm 0.0014$ & $0.481\pm0.003$  \\

\hline\hline                  
\end{tabular}
\tablefoot{Data was obtained from NASA Exoplanet Archive and the Extrasolar Planet Encyclopedia.
}

\end{table*}

\end{document}